*Original article*

**Ramadan Fasting Model Exerts Hepatoprotective, Anti-obesity, and Anti-Hyperlipidemic Effects in an Experimentally-induced Nonalcoholic Fatty Liver in Rats**

**Running title**: Protective effect of Ramadan fasting in NAFLD

Department of Food Science and Nutrition, College of Food and Agriculture Sciences, King Saud University, Riyadh 11451, Saudi Arabia

**Alasmari Abeer A**., MSc, Department of Food Science and Nutrition, College of Food and Agriculture Sciences, King Saud University, Riyadh 11451, Saudi Arabia
**Al-Khalifah Abdulrahman S**., PhD, Department of Food Science and Nutrition, College of Food and Agriculture Sciences, King Saud University, Riyadh 11451, Saudi Arabia
**BaHammam Ahmed S.,** PhD, Department of Medicine, College of Medicine, King Saud University, Riyadh, Saudi Arabia; ashammam@ksu.edu.sa ORCID: 0000-0002-1706-6167
**Alshiban Noura M. S**., MSc., Experimental Surgery and Animal Lab, College of Medicine, King Saud University, Riyadh 11461, Saudi Arabia
**Almnaizel Ahmad T**., MSc., Experimental Surgery and Animal Lab, College of Medicine, King Saud University, Riyadh 11461, Saudi Arabia
**Alodah Hesham S**., PhD, Experimental Surgery and Animal Lab, College of Medicine, King Saud University, Riyadh 11461, Saudi Arabia
**Alhussain Maha H**., PhD, Department of Food Science and Nutrition, College of Food and Agriculture Sciences, King Saud University, Riyadh 11451, Saudi Arabia, ORCID: 0000-0003-0168-6721.

**Correspondence**: Maha H. Alhussain, Department of Food Science and Nutrition, College of Food and Agriculture Sciences, King Saud University, Riyadh 11451, Saudi Arabia, ORCID: 0000-0003-0168-6721; mhussien@ksu.edu.sa

**Abstract**

**Background:**

The epidemic of nonalcoholic fatty liver disease (NAFLD) and its metabolic effects present a serious public health concern. We hypothesized that the Ramadan fasting model (RFM), which involves fasting from dawn to dusk for a month, could provide potential therapeutic benefits and mitigate NAFLD. Accordingly, we aimed to validate this hypothesis using obese male rats.

**Methods:**

Rats were split into two groups (n = 24 per group), and they were given either a standard (S) or high-fat diet (HFD) for 12 weeks. During the last four weeks of the study period, both S- and HFD-fed rats were subdivided into eight groups to assess the effect of RFM with/without training (T) or glucose administration (G) on the lipid profile, liver enzymes, and liver structure (n=6/group).

**Results:**

The HFD+RFM groups exhibited a significantly lower final body weight than that the HFDC group. Serum cholesterol, low-density lipoprotein, and triglyceride levels were significantly lower in the HFD+RFM, HFD+RFM+T, and HFD+RFM+G groups than those in the HFDC group. Compared with the HFD-fed group, all groups had improved serum high-density lipoprotein levels. Furthermore, HFD groups subjected to RFM had reduced serum levels of aspartate transaminase and alanine transaminase compared with those of the HFD-fed group. Moreover, the liver histology has improved in rats subjected to RFM compared with that of HFD-fed rats, which exhibited macro and micro fat droplet accumulation.

**Conclusion:**

RFM can induce positive metabolic changes and improve alterations associated with NAFLD, including weight gain, lipid profile, liver enzymes, and hepatic steatosis.

## INTRODUCTION

Nonalcoholic fatty liver disease (NAFLD) is a serious worldwide health problem. Fatty liver disease has been frequently associated with metabolic dysfunction. To reflect this association more accurately, an international group of experts has suggested changing the name of this condition from NAFLD to metabolic-associated fatty liver disease (MAFLD), which better represents the relationship between fatty liver diseases and metabolic dysfunction based on our current knowledge.[1] In 2016, a meta-analysis of 86 research papers from across 22 countries revealed that the prevalence of NAFLD was 25.2% worldwide.[2] Notably, the Middle East had the highest prevalence of 31.8%, and Asia was ranked second, with a prevalence of 27.4%.[2]

NAFLD is characterized by lipids accumulation in the liver, accompanied by obesity, diabetes mellitus, and/or metabolic syndrome (Met S).[3] The severity of NAFLD can range from a simple accumulation of lipids to more sophisticated phases, including nonalcoholic steatohepatitis (NASH), cirrhosis, and hepatocellular carcinoma.[1] Clinical changes associated with NAFLD in humans and animals include hepatic steatosis, insulin resistance, hyperlipidemia, fasting hyperglycemia, and hepatocyte destruction.[4] Nevertheless, increased *de novo* lipogenesis, gluconeogenesis, oxidative stress responses and inflammation. Conversely, oxidative stress and inflammatory pathways mediate the progression of steatosis to NASH.[5] By 2030, NAFLD is anticipated to be the leading cause of liver transplantation.[6] Hence, to reverse or prevent the detrimental effects of NAFLD, therapeutic and cost-effective strategies that focus on reducing calorie consumption must be adopted, as excessive food intake is a main risk factor for NAFLD and can cause lipid accumulation in the adipose tissue, leading to its expansion.[7] Notably, excessive caloric intake, as well as food consumption throughout the day, can influence liver lipid accumulation.[8] Therefore, abstaining from food during a specific time of the day, such as intermittent fasting may mitigate NAFLD-induced changes.[3]

Interventions, such as calorie restriction and exercise, are effective strategies for NAFLD management.[9] However, given the difficulties in adhering to calorie restriction and its side effects, alternate approaches have been investigated.[5,10] Numerous types of intermittent fasting, including time-restricted feeding, intermittent energy restriction, and Ramadan diurnal intermittent Ramadan fasting have recently gained considerable attention [11], as it reportedly provokes numerous physiological changes that benefit human health. In the fasting period, cells enhance their innate defenses against oxidative and metabolic stress by activating certain pathways. [12]

Ramadan fasting is a unique model of prolonged intermittent fasting, where Muslims abstain from food and drink (even water) consumption from dawn to sunset for one month. Ramadan fasting reportedly contributes to certain physiological changes, including changes in anthropometry (e.g., body weight [BW]) [13], metabolism (e.g., glucose and lipid profile) [14], and immunomodulatory markers. [15,16] Ramadan fasting  may also positively modulate cardiovascular risks, boosting metabolic characteristics by suppressing inflammatory reactions.[17] According to a recent systematic reviews and meta-analyses of research examining the impact of Ramadan fasting on liver function, Ramadan fasting can have a modest yet meaningful positive effect on liver enzymes, potentially offering a temporary safeguard against NAFLD in individuals without pre-existing liver conditions. [3] Nevertheless, not all studies evaluated in the review analyzed baseline characteristics, and a liver biopsy was not performed. Therefore, more rigorous research is required to adjust for potential confounding factors and comprehensively assess the impact of Ramadan fasting on liver function and NAFLD. Additionally, the results of investigations assessing the influence of Ramadan fasting on physiological and immunomodulatory markers were inconsistent.[18] Furthermore, previous studies did not evaluate factors such as lifestyle modifications, dietary changes, and exercise, which could account for the discrepancies observed among different studies. [15]

The health benefit of exercise are advantageous for lowering numerous risk factors and delaying the onset of Met S.[19] A single exercise session can significantly decrease triglyceride (TG) levels and blood pressure, increase high-density cholesterol (HDL), and improve glucose and insulin regulation. Accordingly, combining exercise with fasting might have a positive effect on overall health and reduce the risk of developing NAFLD [20].

Based on the evidence presented, we hypothesized that Ramadan fasting may positively impact the progression of NAFLD. If Ramadan fasting can delay the onset of early-stage NAFLD, further investigation is warranted, particularly among those with Met S, who present an increased risk of developing NAFLD. Intervention studies exploring this hypothesis could provide valuable insights into the potential benefits of Ramadan fasting for preventing or managing NAFLD.

For decades, rats have been used as a dependable and common experimental animal model for human illnesses owing to their physiological similarities to humans. [21] Moreover, several environmental variables can be controlled when performing laboratory assessments in rats. [22] In rat models, chronic feeding with a high-fat diet (HFD) is the most well-documented method for causing obesity and NAFLD in rat models. [23]

We hypothesized that subjecting rats to fasting during the active phase from dusk to dawn for one month could improve the lipid profile, liver enzymes, and liver structure. We called the application of dry fasting during the active phase of the rats as Ramadan fasting Model (RFM) (i.e., the term RFM is specifically proposed for the application of Ramadan fasting on animals in the current scenario).

Therefore, we aimed to investigate the potential health implications of RFM on Wistar albino rats by controlling the dark/light cycle, food intake, and physical activity. By examining these biomarkers and interventions, we aimed to gain novel insights into the potential therapeutic benefits of RFM in mitigating NAFLD.

## MATERIAL AND METHODS

### Experimental animals

Forty-eight Wistar albino male rats (60 ± 10 g; three-week-old) obtained from the Experimental Surgery and Animal Laboratory, College of Medicine, King Saud University were randomly housed individually in ventilated cage racks under controlled conditions (temperature: 22 ± 2°C, humidity: 55 ± 5%). Ethical approval was obtained from the Institutional Animal Care and Use Committee, King Saud University (Ref. No.: KSU-SE-21-66).

### Diets

The present study was performed using standard (S; American Institute of Nutrition AIN-

93G purified Rodent diet) or high-fat diet (HFD; Modified Western Diet with 45 % fat and AIN-93G vitamin and minerals), purchased from Dyets Inc, Bethlehem, PA (catalog numbers: 110700 and 104293; respectively). The S diet provides 2.9 kcal/g (fat constitutes 13% of total energy), the HFD provides 4.73 kcal/g (fat constitutes 45% of total energy).

**Study design**

The total experimental period was 12 weeks. After an adaptation period (one week), rats were divided equally into two main groups. The first group (n=24) was fed the S diet for eight weeks. To induce obesity and NAFLD, the second group (n=24) was fed HFD (45 % fat) *ad libitum* for 8 weeks. After 8 weeks, rats weighing 320 ± 20 g were deemed obese .[23] Subsequently, rats in each group (S group and HFD group) were subdivided into four groups as follows: **(1)** S diet control group (SC) (n=6): rats were fed *ad libitum*. **(2)** S + RFM (S+RFM) (n=6) **(3)** S+ RFM + Training (S+RFM+T) (n=6) ;**(4)** S + RFM + Glucose administration (S+RFM+G) (n=6) ;**(5)** HFD control group (HFDC) (n=6): rats were fed an HFD *ad libitum*; **(6)** HFD + RFM (HFD+RFM) (n=6) **(7)** HFD + RFM + Training (HFD+RFM+T) (n=6) **(8)** HFD+ RFM+ Glucose administration (HFD+RFM+G) (n=6); the duration was four weeks (RFM implementation period).

*RFM protocol*

Given that the normal active phase of rats occurs at night, corresponding to daytime in humans with an 11 h light/13 h dark cycle [24], we implemented the RFM (no food/ beverages) during the active phase. Accordingly, 36 of 48 rats were fasted for four weeks from dusk to dawn (no food or drinks), with approximately 13 h (fasting hours were according to the period from dusk to dawn in the month of the study). Between weeks 9 and 12 (RFM implementation period), food consumption was evaluated by providing rats with precise amounts of food and measuring the quantity left over after 13 h.

### *Glucose administration*

Glucose powder was procured from Sigma Aldrich (MO, USA). At dawn, glucose was orally administered to the S+RFM+G and HFD+RFM+G groups. The glucose dose was selected based on a previous study by Assirey [25]. Glucose was provided at 1.77 mg/kg (diluted with distilled water) to mimic the consumption of three dates to break the fast in humans. Rats had free access to food and water from dusk to dawn.

### *Training session*

The exercise was initiated after familiarization, starting 5–7 days before the formal training. The treadmill (MK-680C No.0807EN03, China) was fixed at 20–25 m/min speed. The exercise protocol was implemented for 10–15 min thrice weekly.[26] A low-grade electric shock at the end of the treadmill forced the rats to run.

### BW and food consumption

BW and food intake were recorded weekly. BW was measured using the EK4150 digital scale (Etekcity, Anaheim, CA, USA). A calibrated scale (0.01 mg precision) was used to measure food consumption. Food intake was calculated as follows: Food consumption (g) = diet provided (g) − diet unconsumed (g) (19).

**Sample collection**

At the end of the experimental study, rats were fasting overnight. Subsequently, the rats were anesthetized using a combination of ketamine/xylazine, administered as a single intraperitoneal injection according to the BW (90 mg/kg ketamine and 10 mg/kg xylazine hydrochloride). Samples of blood were collected by direct cardiac puncture and transferred to a gel-separation yellow cap tube (with no additive), followed by serum separation. The blood was kept at room temperature for approximately 1 h to induce clot formation. To get the serum, the blood was centrifuged at 4000 rpm for 10 minutes. Livers were harvested, weighed, placed on ice, and washed with normal saline. Subsequently, the liver tissues were cut into smaller pieces; then one piece was kept in 10% buffered formalin for histology analysis. The other piece was snap-frozen in liquid nitrogen and maintained at −80°C for further analysis. Animal carcasses were placed in a zip-lock bag and maintained at −20°C until collection by the biohazard company.

**Serum biochemical analyses**

Total cholesterol (CHOL), high-density lipoprotein (HDL), low-density lipoprotein (LDL), and triglycerides (TGs) concentrations were assessed using specific ELISA commercial kits following the kit's instructions (Cat. No. MBS168179; Cat. No. MBS8804648, Cat. No. MBS8806454, and MBS726298 MyBioSource, San Diego, CA, USA, respectively). Serum alanine aminotransferase (ALT) and aspartate aminotransferase (AST) were detected using respective commercial kits, following the manufacturer's instructions (Cat. No. MBS8801684; Cat. No. MBS8804612, MyBioSource, San Diego, CA, USA, respectively).

**Histopathology**

Twenty-four hours after collection, Oil Red O-stain was used to confirm the accumulation of lipid in the liver, as described by Lilie et al. (1943) [27].

**Statistical analysis**

GraphPad Prism 7.1 (GraphPad Software Inc.; San Diego, CA, USA) was used to conduct the statistical analyses. To check for data normality, the Shapiro-Wilk test was applied. The mean and standard deviation (SD) are used to express data. One-way analysis of variance (ANOVA) was used to examine group differences. For post hoc analysis, Tukey's t-test was used to identify statistically significant differences. Statistics were significant at a P-value of 0.05.

## RESULTS

**BW and food intake**

Figure 1-A displays the BW of all groups during the total experimental period (12 weeks). Figures 1-B and 1-C show BW at the end of the experiment for the S groups and HFD groups, respectively. No significant change was observed in BW between the S groups **(Figure 1-B).** However, as shown in Figure 1-C, the rats in the HFD+RFM group had a significantly lower BW than those in the HFDC group (P<0.004). No significant differences were observed among HFD+RFM+T and HFD+RFM+G compared with the HFDC group, although the BW of the HFD group was higher. The food intake did not significantly vary among groups (Table 1).

**Lipid profile**

Figure 2 displays the lipid profile of all groups. No significant variations were detected between S groups in serum CHOL and LDL levels (**Figure 2-A and 2-C,** respectively). The S+RFM, S+RFM+T, and S+RFM+G groups had greater serum HDL levels than that in the SC group. The serum level of HDL was significantly higher in S+RFM+T compared with SC (P=0.012) and S+RFM+G (P=0.047) **(Figure 2-E).**

The TG level of the SC groups was significantly lower than in the S+RFM+G group (P<0.0009). In addition, the level of TG in the S+RFM+G was significantly greater than S+RFM group (P<0.0003) (**Figure 2-G**).

In terms of HFD groups, the HFDC group had a greater lipid profile, including serum CHOL and LDL levels, than the other groups (**Figure 2-B and 2-D,** respectively**)**. The HFD+RFM, HFD+RFM+T, and HFD+RFM+G groups exhibited significantly lower CHOL levels than the HFDC group (P<0.0001) **(Figure 2-B).** In addition, the level of serum LDL was significantly lower in the HFD+RFM+T group than that in the HFD+RFM (P=0.002) **(Figure 2-D).**

The HFDC group had a significantly lower HDL level than that in the HFD+RFM (P<0.0001) and HFD+RFM+T (P<0.0001) groups. Moreover, the serum HDL level was significantly lower in the HFD+RFM+G compared with HFD+RFM (P<0.0001) and HFD+RFM+T (P<0.0001) groups (**Figure 2-F**).

The HFDC group exhibited greater serum TG levels than that in the other groups. Furthermore, TG levels of the HFD+RFM (P=0.029) and HFD+RFM+T (P=0.031) groups were significantly lower than those of the HFD+RFM+G group (**Figure 2-H**).

**Liver enzymes:**

Figure 3 illustrates the liver enzymes of all groups. serum level of ALT was significantly greater in S+RFM+T than in the SC group (P=0.0002). serum level of ALT was significantly higher in S+RFM+T than S+RFM (P=0.001) and S+RFM+G group (P=0.018) **(Figure 3-A).** Serum levels of AST did not significantly differ in the S+RFM, S+RFM+T, and S+RFM+G groups compared with that in the SC group **(Figure 3-C).** The HFDC group had a greater ALT and AST serum level than the other groups **(Figure 3-B, 3-D)**. In the HFD groups, the serum ALT level was significantly lower in the HFD+RFM (P=0.0005) and HFD+RFM+T (P=0.0002) groups than those in the HFDC group. The serum levels of ALT were significantly higher in HFD+RFM+G than in HFD+RFM and HFD+RFM+T (P=0.013) and (P=0.004), respectively **(Figure 3-B).** The serum levels of AST were significantly lower in the HFD+RFM (P<0.0001), HFD+RFM+T(P<0.0001), and HFD+RFM+G (p=0.001) groups compared with the HFDC group **(Figure 3-D).** The serum levels of AST were significantly higher in HFD+RFM+G than in HFD+RFM and HFD+RFM+T (P=0.013) and (P=0.004), respectively **(Figure 3-D)**.

**Hepatic steatosis**

Liver histology is illustrated in Figure 4. The SC, S+RFM, S+RFM+T, and

S+RFM+G groups displayed normal liver histology with no lipid accumulation in Figures 4 (A, B, C, D). Oil Red O-stained showed hepatic lipid accumulation in the HFDC group (**Figure 4-E**). In contrast, the HFD+RFM+T groups showed reduced fat vacuole deposition in the hepatocytes compared with the HFDC group (**Figures 4-G and 4-E**). Notably, the HFD+RFM and HFD+RFM+G groups exhibited an improvement in lipid droplets compared with HFDC groups (**Figure 4-F and 4-H**).

**Liver Weights:**

Figure 4 displays the liver weights among all groups. No significant differences were observed among SC, S+RFM, S+RFM+T, and S+RFM+G in liver weights (**Figure 4-I**). The rats in the HFDC group had greater liver weights than other groups. The liver weight of the HFDC group was significantly higher than HFD+RFM+T (P=0.0005). Moreover, liver weight was significantly lower in HFD+RFM+T compared with HFD+RFM+G (P=0.0047) **(Figure 4-J)**.

## DISCUSSION

In the present study, we examined the health implications of RFM on NAFLD and certain accompanying alterations, namely BW, lipid profile, and liver enzyme levels in HFD-induced obese rats while controlling the dark/light cycle, food intake, and physical activity. Our findings revealed that RFM (rats were fasted during their active phase, from dusk to dawn) could reduce BW, ameliorate lipid profile and liver enzymes, and attenuate hepatic steatosis.

Obesity and related metabolic disorders are the most well-known risk factors for NAFLD and NASH. Chronic HFD administration can induce dyslipidemia and hepatic steatosis via several mechanisms. The influx of free fatty acid post-HFD administration can stimulate lipogenesis and increase fat droplets in the liver. Thus, chronic HFD feeding is deemed the best strategy for inducing obesity, metabolic abnormalities, and NAFLD in rats.[23] Concordantly, we used this model to mimic the human metabolism when chronically fed with HFD.

Considering the present study, one important observation is the improvement in hepatic steatosis in the HFD rats subjected to RFM, indicating a potential hepatoprotective effect that could be related to improved lipids profiles and fatty acid oxidation. Furthermore, RFM was associated with. No previous report has systematically assessed the impact of Ramadan fasting on liver histology; however, other the impact of experimental fasting models has been examined in animal models. Marinho et al. [28] reported that intermittent fasting protocol, i.e., alternating 24-h feeding and 24-h fasting periods, could enhance liver function enzymes, liver steatosis, and inflammation in obese mice subjected to an obesogenic and pro-inflammatory diet. Furthermore, Kord Varkaneh et al. tested the effect of 5:2 intermittent fasting (20–25% restriction of energy for two nonconsecutive days/week and *ad libitum* feeding for the rest of the days) NAFLD patients compared with a control group. The authors observed that intermittent fasting reduces the risk of NAFLD by attenuating hepatic steatosis, enzymes, and lipid profiles. [29] NAFLD is reportedly linked to insulin resistance and oxidative stress, which may be alleviated by intermittent fasting, including Ramadan fasting [30]. Moreover, intermittent fasting has been associated with suppressed lipolysis, improved fatty acid oxidation, and the biogenesis of mitochondria in the hepatocyte. [3]

Herein, we observed that RFM diminished liver enzymes in HFD rats, possibly via a diet-dependent action on AST and ALT. Similarly, a recent meta-analysis assessing six studies (three studies of Ramadan fasting and the other three of experimental fasting) has revealed that intermittent fasting could diminish the risk of NAFLD in humans by reducing the level of hepatic enzymes.[31] Results from human intervention trials have also indicated that Ramadan fasting can reduce liver enzyme levels. For instance, a meta-analysis involving 20 cohort studies suggested that Ramadan fasting could improve liver enzymes [3]. The authors proposed that the beneficial effects of Ramadan fasting on hepatic enzymes among healthy participants could benefit patients with NAFLD by improving the liver enzyme levels, warranting further exploration in future investigations. [3, 32] The positive effects of Ramadan fasting are frequently attributed to a reduction in BW and/or body fat. Moreover, the possible mechanisms underlying the Ramadan fasting induced liver benefits could be related to the impact on circadian biology, gut flora, and modifiable lifestyle choices [33,34]. Intermittent fasting is substantially more complex than a simple cycle of feeding and fasting. Intermittent fasting can alter the gut microbiome composition and boost the immune system.[34] Furthermore, shifting meal time from day to night during Ramadan fasting could alter the circadian rhythm.[35] As mentioned previously, these changes are critical for regulating physiological functions [36]. The impact of intermittent fasting on lipid oxidation is independent and possibly mediated by regulating fatty acid oxidation-related genes (i.e., Sirtuin 1 (SIRT1) and adenosine monophosphate-activated protein kinase (AMPK) and their downstream target genes. [37] However, in the present study, we did not assess these genes; therefore, further studies are required to ascertain the possible mechanisms of RFM on NAFLD by assessing gene expression of SIRT1 and AMPK. intermittent fasting also potentially activates AMPK expression, which increases the oxidation of fatty acid and suppresses *de novo* lipogenesis in the liver. This may be another possible mechanism through which RFM ameliorates liver steatosis. [37]

We observed that intervention with RFM significantly alleviated the metabolic status of HFD rats and significantly lowered BW and liver steatosis, indicating the anti-obesity and anti-NAFLD potential of RFM. Although there was no substantial variation in food intake, Intermittent fasting positively impacts metabolism and reduction in BW. [38,39] Moreover, these findings highlight the ability of RFM to utilize fat in adipose tissue, the main approach used to reduce BW. Similar to our data, a recent systematic review investigated the effect of Ramadan fasting on the BW of healthy participants and reported a significant yet small reduction in BW. [40] Moreover, a systematic review and meta-analysis of data gathered from 70 articles [41] reported that Ramadan fasting could significantly decrease the fat percentage in the overweight/obese but not in the normal weight groups. While the findings of some studies are in agreement with our findings [14], others failed to detect appreciable changes in BW after Ramadan fasting. [42] Notably, all studies evaluated in the aforementioned systematic reviews were conducted in the free-living environment without controlling for lifestyles, such as dietary habits, circadian rhythm, and exercise, which can influence BW. [3,31] Accordingly, the observed discrepancies in BW results may be attributed to a free-living environment without regulating potential confounders and disruption in the lifestyle. [15]

The main characteristics of NAFLD, which are frequently observed in HFD-experimental animals, include hyperlipidemia, increased hepatic *de novo* lipogenesis, and steatosis. [43] Furthermore, inflammation and oxidative stress are major hallmarks of NAFLD. Concordantly, in the present study, HFD rats exhibited typical features of hyperlipidemia, including elevated levels of TGs, CHOL, and LDL and the development of liver steatosis. Notably, we observed that RFM/RFM+T intervention could lower the levels of CHOL, LDL, and TGs in HFD-fed rats. In addition, the HDL level significantly improved in lean and obese rats, indicating the substantial benefit of RFM. Several lines of evidence have reported that RFM can attenuate the levels of these lipid parameters. For example, a meta-analysis assessing 85 studies. [44] revealed that RF can modestly improve TG and HDL in healthy individuals, consistent with our findings. [44] A recent meta-analysis included approximately 101 prospective papers that reported that Ramadan fasting positively impacts Met S components (HDL, TG, blood pressure, and waist circumference).[45] Moreover, Temizhan et al. conducted time-restricted feeding on adults for four weeks and documented a substantial reduction in CHOL, LDL, and TGs levels, whereas HDL levels were increased from baseline. [46] Conversely, some studies have documented the lack of significant changes in CHOL and TG levels. [47-49] The small sample size and increased dessert consumption during Ramadan may have contributed to this discrepancy.

Notably, the reduced BW could explain the mechanism by which RFM improves lipid profiles. [16] Recent research has revealed the mechanisms of action through which intermittent fasting may improve lipid profiles. Peroxisome proliferator-activated receptor (PPAR) and peroxisome proliferator-activated receptor coactivator 1 (PGC-1) are nuclear expressions in the liver. PPAR and PGC-1 promote fatty acid oxidation and apolipoprotein A production while reducing apolipoprotein B synthesis. Moreover, hepatic TG and very LDL (VLDL) synthesis reportedly decrease while fatty acid oxidation increases. An overall reduction in serum VLDL and LDL levels may occur owing to these physiologic changes. [50] Nevertheless, a comprehensive investigation is required to evaluate the effect of RFM on PPAR and PGC-1 at molecular levels in individuals with NAFLD.

The present study has several important limitations that should be acknowledged. Firstly, we did not measure adipose tissue weight, muscle tissue weight, and liver triglyceride (TG) levels, which are critical parameters for a comprehensive understanding of the observed effects. It is strongly recommended that future research incorporates these measurements to provide a more thorough assessment. Additionally, further well-designed studies are warranted to elucidate the precise molecular mechanisms underlying the benefits of RFM. Techniques such as immunohistochemistry and investigations into hepatic glucose and lipid regulators would be valuable in unraveling the underlying biological processes. These approaches would contribute to a deeper understanding of the mechanisms involved and strengthen the scientific basis for the observed outcomes.

In conclusion, our data present the first liver biopsy evidence demonstrating the beneficial effects of RFM on NAFLD in a rat model with obesity, suggesting that RFM exerts potential hepatoprotective, hypolipidemic effects. In HFD-induced obese rats, RFM can facilitate weight loss and improve the lipid profile and hepatic enzyme levels. Moreover, RFM alone was valuable in inducing weight loss, hypolipidemia, and alleviating hepatic steatosis.

# Tables

**Table 1:** Food intake of rats fed the standard and high-fat diet (HFD).

| Groups | Food intake (G/100 g BW) |
|--------|--------------------------|
| SC | 11.5 ± 2.8 |
| HFD | 12.3 ± 3.1 |
| S+RFM | 11.8 ± 1.8 |
| HFD+RFM | 11.9 ± 1.2 |
| S+RFM+T | 9.8 ± 1.1 |
| HFD+RFM+T | 10.3 ± 1.1 |
| S+RFM+G | 12.3 ± 2.1 |
| HFD+RFM+G | 13.5 ± 2.5 |
| **P-value** | P=0.108 |

Data are expressed as means ± standard deviation (SD) (n = 6 rats/group). BW, body weight; SC, control; S, standard diet; HFD, high-fat diet; RFM, Ramadan Fasting Model; T, training; G, Glucose administration.

## Figures Legends

**Figure 1**:

Figure 1-A: Presents the weight over time from the beginning of the study until its conclusion in all groups.

Figure 1-B: It illustrates the body weight in all standard groups. One-way ANOVA revealed no significant differences.

Figure 1-C: It illustrates the body weight in all HFD groups. One-way revealed significant differences between HFD and HFD+RFM.....

Data in (figure B) are expressed as mean ± standard deviation (SD) (n = 6). Data were tested by one-way ANOVA followed by Tukey's t-test for post hoc analysis. The bracket represents statistically significant differences among groups at **P-value ≤ 0.05; **P-value ≤ 0.005; ***P-value ≤ 0.0005. SC, S diet control; S, Standard diet; HFD, high-fat diet; RFM, Ramadan fasting Model; T, training; G, Glucose administration.

**Figure 2:** Lipid profiles of all experimental rat groups.

Figure 2-A: Shows the Serum levels of cholesterol (CHOL) in all standard groups. One-way ANOVA revealed no differences.

Figure 2-B: Shows the Serum levels of cholesterol (CHOL) in all HFD groups. One-way ANOVA revealed significant differences between HFD, HFD+RFM, HFD+RFM+T, and HFD+RFM+G.

Figure 2-C: Shows the Serum levels of low-density lipoproteins (LDL) in all standard groups. One-way ANOVA revealed no differences.

Figure 2-D: Shows the Serum levels of low-density lipoproteins (LDL) in all HFD groups. One-way ANOVA revealed a significant differences between HFD , HFD+RFM, HFD+RFM+T, and HFD+RFM+G. also shows a significant differences between HFD+RFM and HFD+RFM+T

Figure 2-E: Shows the Serum levels of high-density lipoproteins (HDL) in all standard groups. One-way ANOVA revealed significant differences between SC and S+RFM+T and also between S+RFM+T and S+RFM+G.

Figure 2-F: Shows the Serum levels of High-density lipoproteins (HDL) in all HFD groups. One-way ANOVA revealed significant differences between HFD, HFD+RFM, HFD+RFM+T, and HFD+RFM+G. It also shows significant differences between HFD+RFM and HFD+RFM+T.

Figure 2-G: Shows the Serum levels of triglycerides (TGs) in all standard groups. One-way ANOVA revealed significant difference between SC and S+RFM+G and also between S+RFM and S+RFM+G.

Figure 2-H: Shows the Serum levels of triglycerides (TGs) in all HFD groups. One-way ANOVA revealed significant differences between HFD, HFD+RFM, HFD+RFM+T, and HFD+RFM+G. Also shows significant differences between HFD+RFM+G and other HFD groups.

Data are expressed as mean ± standard deviation (SD) (n = 6)**.** Data in each diet group were tested by one-way ANOVA followed by Tukey's t-test for post hoc analysis**.** The bracket represents statistically significant differences among groups at *P-value ≤ 0.05; **P-value ≤ 0.005; ***P-value ≤ 0.0005. SC, S diet control; S, Standard diet; HFD, high-fat diet; RFM, Ramadan fasting Model; T, training; G, Glucose administration.

**Figure 3:** Liver enzyme levels in all experimental rat groups.

Figure 3-A: Shows the Serum levels of alanine aminotransferase (ALT) in all standard groups. One-way ANOVA revealed significant differences between SC and S+RFM+T and S+RFM, S+RFM+T, and S+RFM+G.

Figure 3-B: Shows the Serum levels of alanine aminotransferase (ALT) in all HFD groups. One-way ANOVA revealed significant differences between HFD, HFD+RFM, HFD+RFM+T, and HFD+RFM+G.

Figure 3-C: Shows the Serum levels of aspartate aminotransferase (AST) in all standard groups. One-way ANOVA revealed no significant differences between groups.

Figure 3-D: Shows the Serum levels of alanine aminotransferase (AST) in all HFD groups. One-way ANOVA revealed significant differences between HFD, HFD+RFM, HFD+RFM+T, and HFD+RFM+G. Also presented significant differences between HFD+RFM+G, HFD+RFM, and HFD+RFM+G

Data are expressed as mean ± standard deviation (SD) (n = 6)**.** Data in each diet group were tested by one-way ANOVA followed by Tukey's t-test for post hoc analysis**.** The bracket represents statistically significant differences among groups at *P-value ≤ 0.05; **P-value ≤ 0.005; ***P-value ≤ 0.0005. SC, S diet control; S, Standard diet; HFD, high-fat diet; RFM, Ramadan fasting Model; T, training; G, Glucose administration.

**Figure 4A-H:** Oil Red O-stained images of livers from all experimental rat groups and Liver weights. Sections (A, B, C, D) represent SC, S+RFM, S+RFM+T, and S+RFM+G groups, respectively, displaying normal liver tissue free of lipid deposition, given the absence of significant red color. Section (E) is a representative image of HFD rats exhibiting lipid accumulation, given the appearance of an intensive focal red spot. Sections (D, H) represent HFD+RFM and HFD+RFM+G groups, showing a slight reduction in lipid accumulation in liver tissues. Section (F) represents the HFD+RFM+T group and shows a better reduction in lipid accumulation in liver tissue than (D and H) groups, given the appearance of tissue color. Oil Red O staining. Scale bar = 400 μm, 100×. SC, S diet control; S, Standard diet; HFD, high-fat diet; RFM, Ramadan fasting Model; T, training; G, Glucose administration.

Figure 4-I: Shows the Liver weights in all standard groups. One-way ANOVA revealed no differences.

Figure 4-J: Shows the Liver weights in all standard groups. One-way ANOVA revealed significant differences in HFD+RFM+T compared with HFD, and HFD+RFM+G.

Data are expressed as mean ± standard deviation (SD) (n = 6). Data in each diet group were tested by one-way ANOVA followed by Tukey's t-test for post hoc analysis. The bracket represents statistically significant differences among groups at *P-value ≤ 0.05; **P-value ≤ 0.005; ***P-value ≤ 0.0005. SC, S diet control; S, Standard diet; HFD, high-fat diet; RFM, Ramadan fasting Model; T, training; G, Glucose administration.

**Figure 1:**

**A**

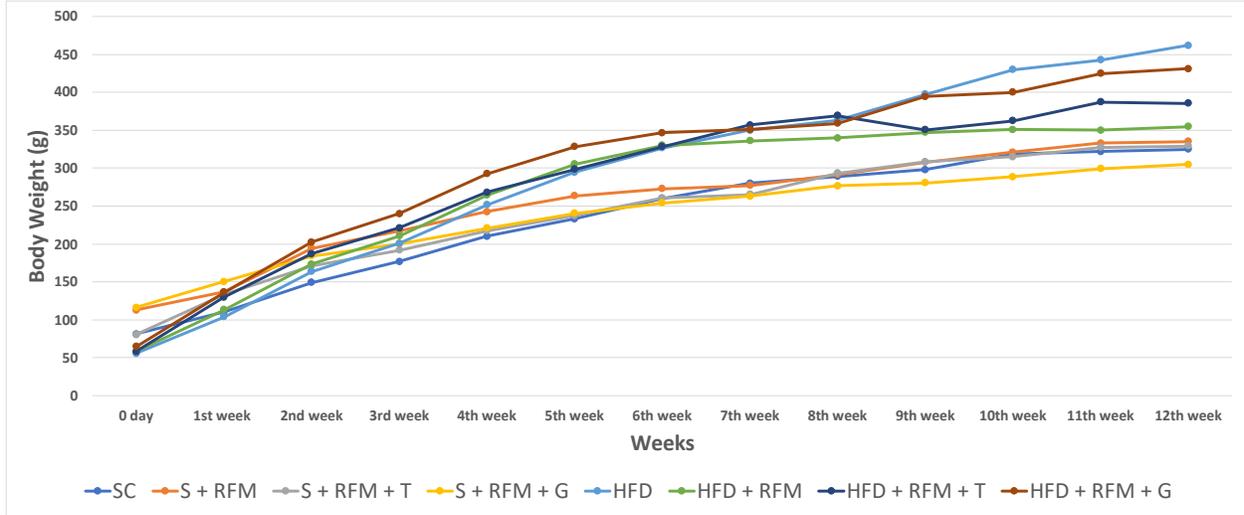

**B**

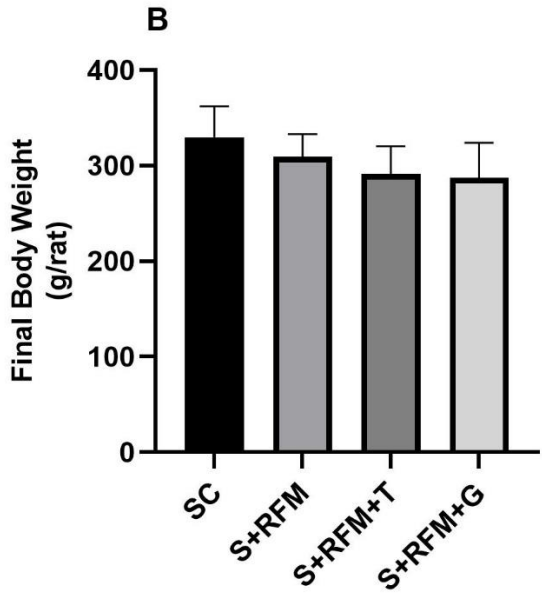

**C**

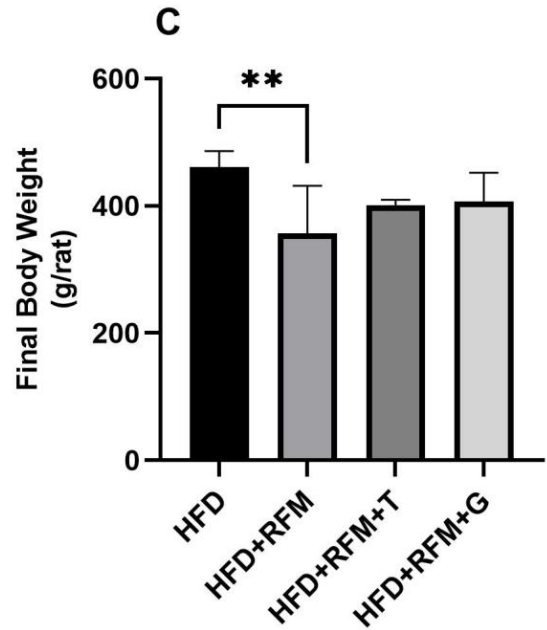

**Figure 2:**

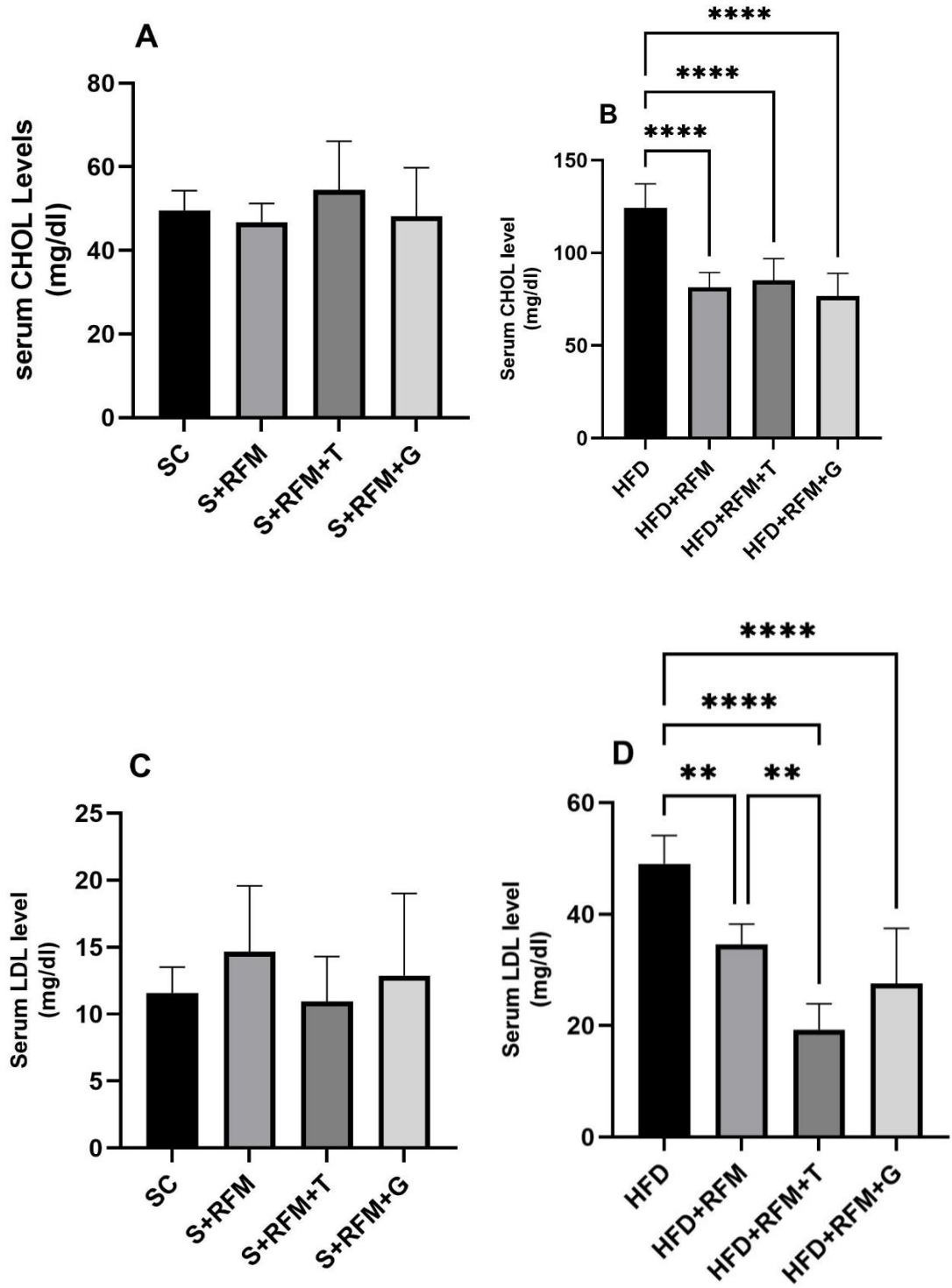

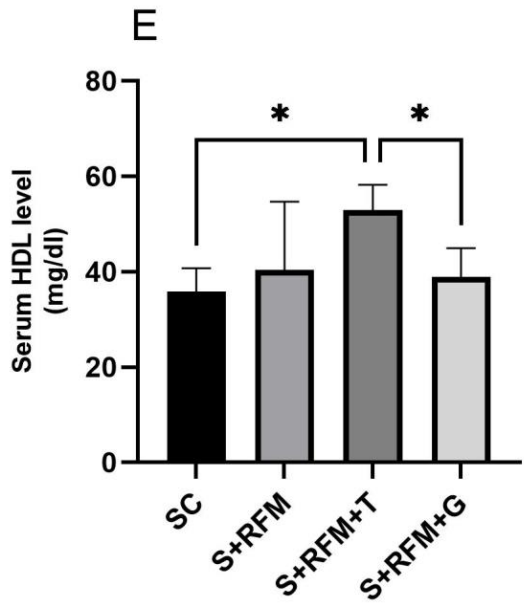

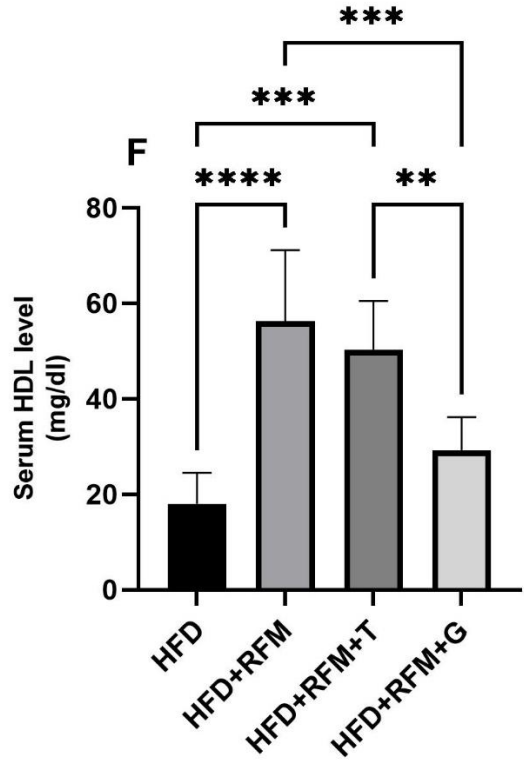

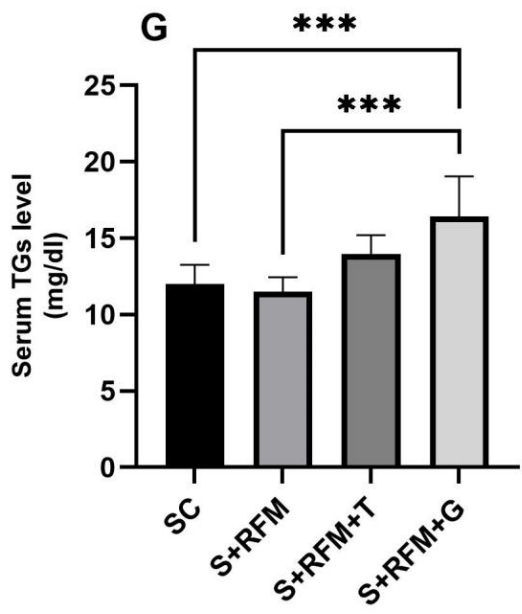

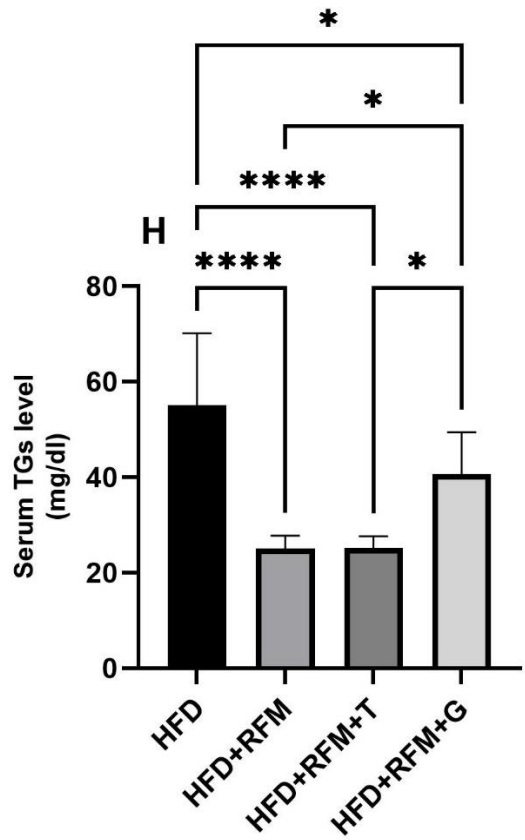

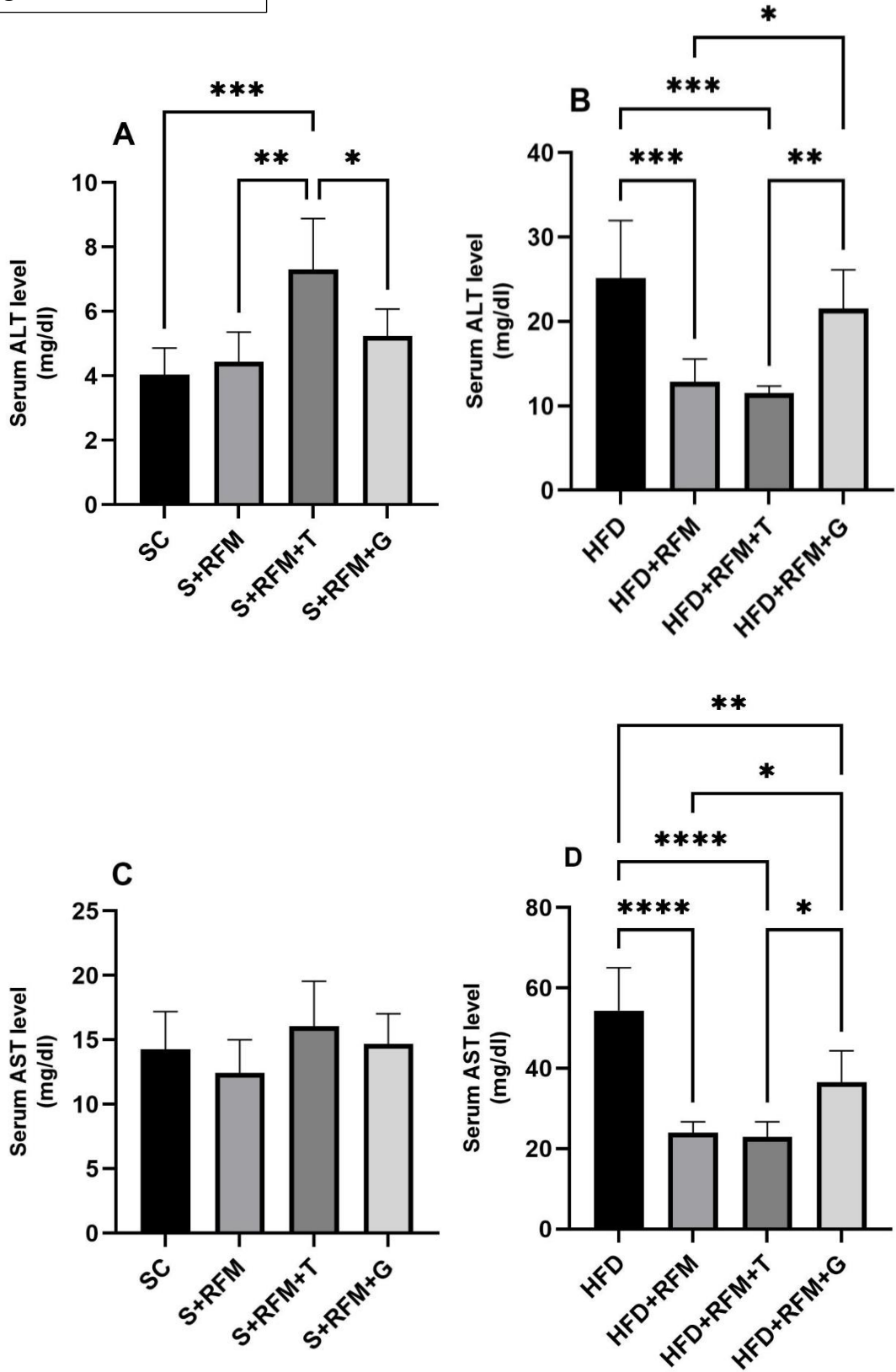

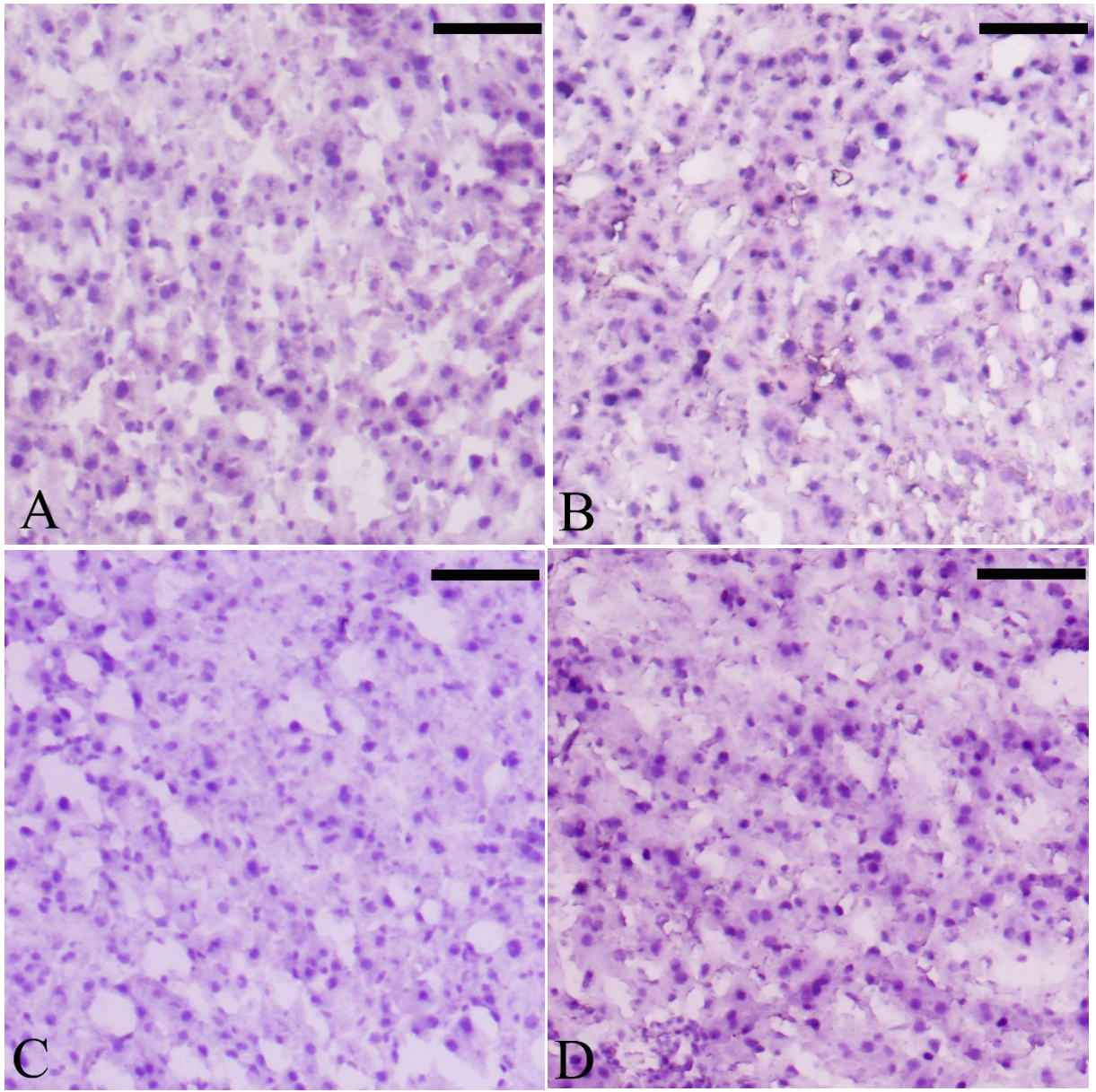

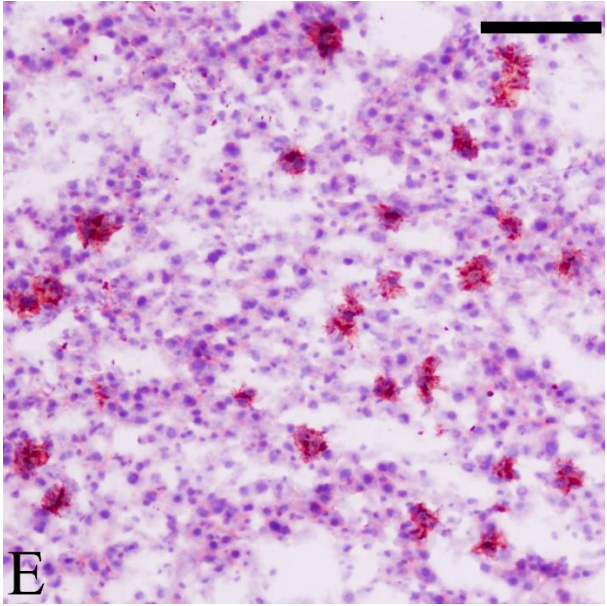
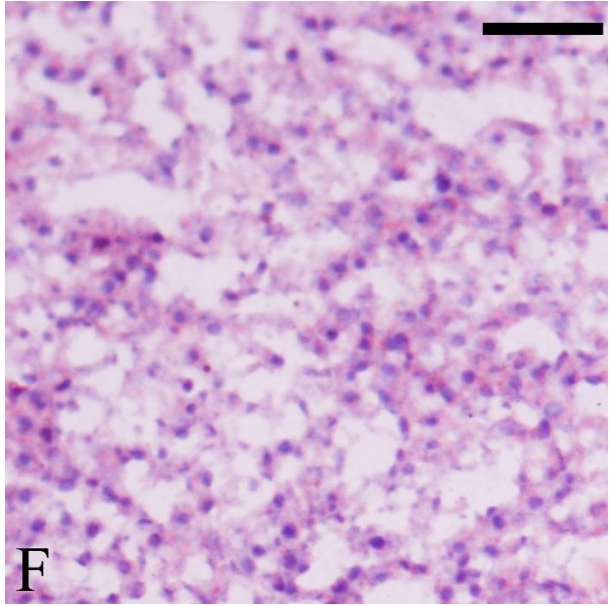
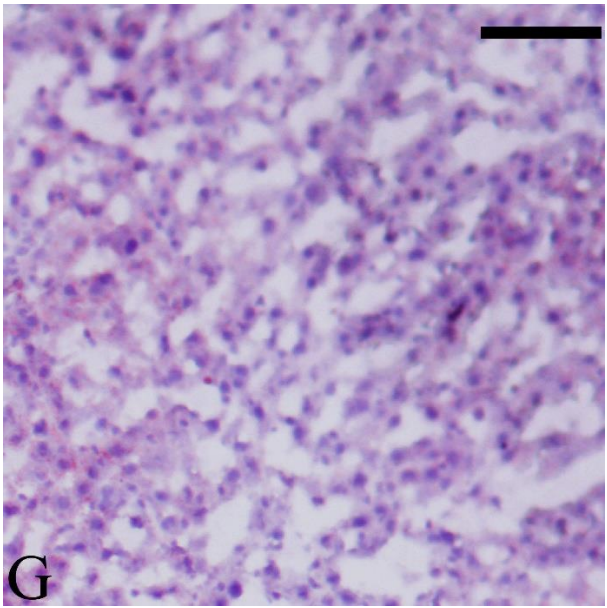
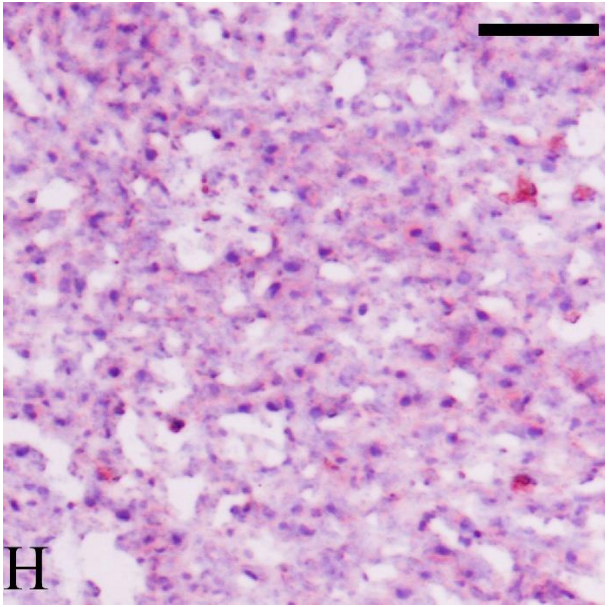

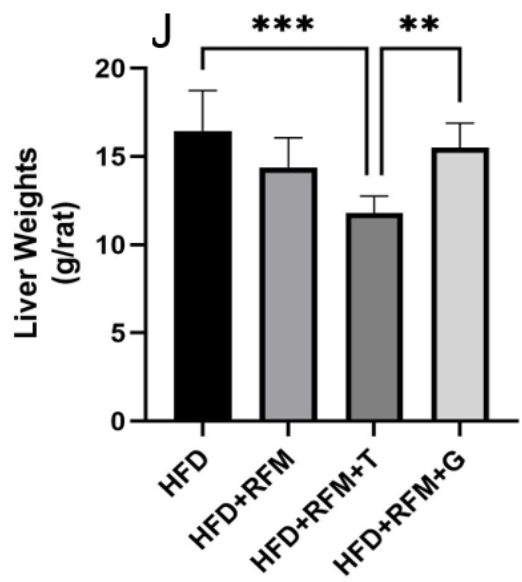

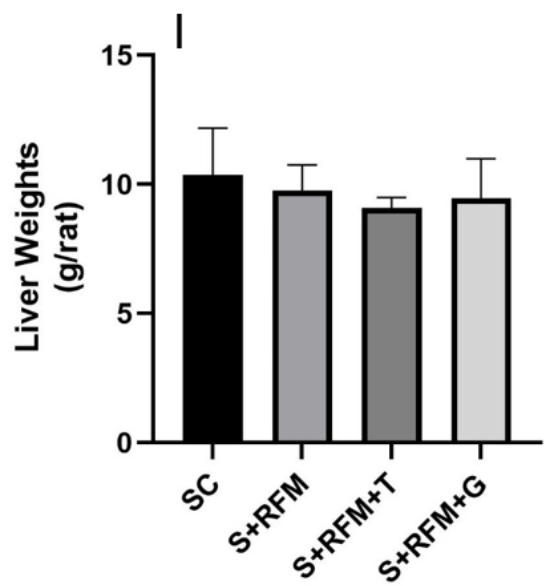